\documentclass[%
 reprint,
 amsmath,amssymb,
 aps,
]{revtex4-1}
\usepackage{graphics}
\usepackage{graphicx}
\usepackage{dcolumn}
\usepackage{bm}

\begin{document}

\preprint{APS/123-QED}

\title{Stringent Statistical Fluctuation Analysis for Quantum Key Distribution Considering After-pulse Contributions
 }
\author{Hongxin Li}
\email{lihongxin830@163.com}
\affiliation{State Key Laboratory of Mathematical Engineering and Advanced Computing, Zhengzhou, Henan, China }
\affiliation{University of Foreign Languages, Luoyang, Henan, China }
\author{Haodong Jiang}
\affiliation{State Key Laboratory of Mathematical Engineering and Advanced Computing, Zhengzhou, Henan, China }
\author{Ming Gao}
\email{gaoming@nudt.edu.cn}
\affiliation{State Key Laboratory of Mathematical Engineering and Advanced Computing, Zhengzhou, Henan, China }
\author{Zhi Ma}
\email{Ma\_zhi@163.com}
\affiliation{State Key Laboratory of Mathematical Engineering and Advanced Computing, Zhengzhou, Henan, China }
\author{Chuangui Ma}
\affiliation{State Key Laboratory of Mathematical Engineering and Advanced Computing, Zhengzhou, Henan, China }
\author{Wei Wang}
\affiliation{University of Foreign Languages, Luoyang, Henan, China }

\date{\today}

\begin{abstract}
Statistical fluctuation problems are faced by all quantum key distribution (QKD) protocols under finite-key condition. Most of the current statistical fluctuation analysis methods work based on independent random samples, however, the precondition can¡¯t be always satisfied on account of different choice of samples and actual parameters. As a result, proper statistical fluctuation methods are required to figure out this problem. Taking the after-pulse contributions into consideration, we give the expression of secure key rate and the mathematical model for statistical fluctuations, focusing on a decoy-state QKD protocol (Sci Rep. 3, 2453, 2013) with biased basis choice. On this basis, a classified analysis of statistical fluctuation is represented according to the mutual relationship between random samples. First for independent identical relations, we make a deviation comparison between law of large numbers and standard error analysis. Secondly, we give a sufficient condition that Chernoff bound achieves a better result than Hoeffding¡¯s inequality based on independent relations only. Thirdly, by constructing the proper martingale, for the first time we represent a stringent way to deal with statistical fluctuation issues upon dependent ones through making use of Azuma¡¯s inequality. In numerical optimization, we show the impact on secure key rate, the ones and respective deviations under various kinds of statistical fluctuation analyses.
\begin{description}
\item[PACS numbers]
03.67.Dd, 42.81.Gs, 03.67.Hk
\end{description}
\end{abstract}

\pacs{Valid PACS appear here}
\maketitle


\section{\label{sec:level1}Introduction}
Quantum key distribution (QKD) protocols \cite{bennett1984quantum,ekert1991quantum,gisin2002quantum} can provide unconditional security for telecommunications based on the physical law of quantum mechanics \cite{mayers2001unconditional,bennett1996limitations,lo1999unconditional,shor2000simple,renner2005security}.
However, in real-life QKD systems, there is a large gap between the ideal conditions and the practical ones.
In order to figure out the problem, Gottesman et al. \cite{GLLP2004security} gave an intensive study aiming at the real system imperfections, and provided a key rate computation formula called GLLP.
Based on GLLP formula, decoy-state protocols \cite{hwang2003quantum,wang2005beating,lo2005decoy,ma2005security} were proposed so as to improve secure key rate and detect photon-number-splitting (PNS) attack\cite{brassard2000limitations,pns2002quantum}. For the first 20 years since the proposal of BB84 protocol \cite{bennett1984quantum}, QKD postprocessing procedure had been mainly analyzed in the asymptotic assumption that the length of transmission code is infinite.
Nevertheless, in real-world situations, the two parties often communicate within a limited time.
As a result, many efforts \cite{mayers2001unconditional,ma2005security,hayashi2007upper,scarani2008quantum,cai2009finite,tomamichel2012tight,ma2012statistical,wei2013decoy,zhou2014tightened,zhou2015making,lucamarini2013efficient,lim2014concise,curty2014finite} have been spent on the finite-key effect in QKD postprocessing schemes.

The security of QKD under finite-key condition was first considered by Mayers \cite{mayers2001unconditional}, and soon afterwards, by using standard error analysis, Ma \cite{ma2005security} gave a detailed analysis concerning five kinds of statistical fluctuations.
Based on the work in \cite{ma2005security,lo2008efficient}, Wei et al. \cite{wei2013decoy} proposed an efficient decoy-state QKD protocol with biased basis choice and adopted standard error analysis to tackle statistical fluctuation problems.
More recently, Zhou et al. \cite{zhou2015making} proposed an improved statistical fluctuation method in measurement-device-independent (MDI) QKD also based on standard error analysis, which raises the key rate of the recent Shanghai experimental set-up \cite{tang2014mdi} by more than 600 times.

Under universal composable (UC) \cite{canetti2001uc} security which is accepted and considered as the most stringent security standard in classical and quantum fields, Scarani and Renner \cite{cai2009finite} presented the finite-key QKD security analysis using Law of large numbers \cite{scarani2008quantum,cai2009finite,Cover2006elements}.
By adopting Law of large numbers to deal with statistical fluctuation issues, Cai and Scarani \cite{cai2009finite} discussed the secure key generation rate in their decoy-state protocol and obtain the shortest sending code length for generating secure key.
Based on their research, Lucamarini et al. \cite{lucamarini2013efficient} used beta distribution and CP test \cite{clopper1934confidence} to handle the finite-key effects and gave an experimental verification for their decoy-state QKD protocol.

By taking advantage of uncertainty relationship for smooth entropies, Tomamichel et al. \cite{tomamichel2012tight} provided an upper bound of the final key.
Based on their work, through utilizing Hoeffding's inequality \cite{Hoeffding1963probability} to analyze statistical fluctuations in decoy-state QKD protocol, Ref. \cite{lim2014concise}  achieved a good estimation result.
Since then large deviation theories \cite{Hoeffding1963probability,chernoff1952measure} have attracted lots of attentions in dealing with statistical fluctuation problems.
Applying Chernoff bound \cite{curty2014finite,chernoff1952measure} to perform parameter estimation, Curty et al. demonstrated the feasibility of long-distance implementations of MDIQKD in finite-code length.

In most of the above QKD postprocessing schemes lim2014concise \cite{ma2005security,cai2009finite,wei2013decoy,lim2014concise,grimmett2001brobability,azuma1967dependent,curty2014finite} , the main precondition of statistical fluctuation analysis is almost the same, which requires an independent mutual relation between random samples. Renner \cite{renner2007symmetry} pointed out that symmetry of large physical systems implies independence of subsystems, thus the independent relationship is reasonable as long as the system¡¯s physical properties are robust under small disturbances. However, when it comes to after-pulse contributions, the detection event might be caused by the after-pulse of the prior one. Many efforts have been spent on investigating after-pulse \cite{nambu2011afterpulse,nato2009afterpulse} and its effect \cite{rolando2013pns,akio2002method}, concentrating on the correlations of neighbouring pulses and the security analysis. Somma et al.\cite{rolando2013pns} demonstrated that if the independent condition is broken by after-pulse contributions, then the security parameters of decoy-state implementations could be worse, and they also pointed out that the eavesdropper possesses the ability to correlate her PNS attack which breaks the independent condition, so that the corresponding security analysis would be invalid in the above mentioned protocols. Furthermore, the after-pulse effect will lower the secure key rate, which is even worse at higher gate frequencies because of a large number of after-pulse-related errors\cite{akio2002method}. As a result, the after-pulse contributions can¡¯t be neglected, and we need to find a stringent way to reconsider the security analysis, especially statistical fluctuation issues based on dependent random samples, which is the motivation behind our analysis.

In this paper, we focus on the mathematical models, classifications and deviation comparisons for various kinds of statistical fluctuation analyses, including standard error analysis, Law of large numbers, Hoeffding¡¯s inequality and Chernoff bound. However, these methods are not the accurate ways to deal with statistical fluctuation issues when taking the after-pulse contribution into consideration. In order to solve the problem, for the first time we provide a new way called Azuma¡¯s inequality \cite{azuma1967dependent} to deal with dependent random samples.
Since the efficiency of a practical QKD system needs to care about the total quantum resource consumption, it is important to take the sending qubits as the total random samples.

The article is organized as follows. Section \ref{sec:level2} gives an analysis of after-pulse contributions based on Wei¡¯s decoy- state protocol \cite{wei2013decoy}. \ref{sec:level3}, we set up mathematical models for standard error analysis and Law of large numbers based on independent identically distributed (i.i.d.) random samples, and present a deviation comparison. For independent random samples only, we give a similar analysis for large deviation theories in section \ref{sec:level4}. In section \ref{sec:level5}, we bring a stringent way to solve the statistical fluctuation issues on dependent samples. Simulations for secure key rates and respective deviations of the five kinds of estimation methods are given in section \ref{sec:level6}. At last, some conclusions are presented in section \ref{sec:level7}.

\section{\label{sec:level2} Analysis of after-pulse contributions}
The after-pulse effect is due to carrier traps, and the uncontrollable clicks will contribute to the quantum bit error rate(QBER). To make a detailed analysis, here we choose vacuum+weak decoy-state QKD protocol with biased basis choice \cite{wei2013decoy}. Taking after-pulse contributions into consideration, we need to reconsider the expected detection rate, the yield of vaccum decoy-state yield and the observed QBER.

\subsection{\label{sec:level2A} Parameter description}
Denoting \({p_{ap}}\) as the after-pulse probability, here we suppose \({p_{ap}} = 4 \times {10^{ - 2}}\)\cite{lim2014concise}. Then the parameters mentioned above change their expressions as follows.
\begin{equation}
\begin{array}{l}
\begin{array}{ccl}
\displaystyle {D_\mu } &=& {Q_\mu }\left( {1 + {p_{ap}}} \right) = \left( {1 - {e^{ - \eta \mu }}\left( {1 - {Y_0}} \right)} \right)\left( {1 + {p_{ap}}} \right)\\
\displaystyle {D_\nu } &=& {Q_v}\left( {1 + {p_{ap}}} \right) = \left( {1 - {e^{ - \eta v}}\left( {1 - {Y_0}} \right)} \right)\left( {1 + {p_{ap}}} \right)\\
\displaystyle {Y_0}' &=& 2{p_{dc}}\left( {1 + {p_{ap}}} \right)\\
\displaystyle {E_\mu }' &=& \frac{{\left( {{e_0}{Y_0}' + {e_d}(1 - {e^{\eta \mu }})(1 - {Y_0}') + {p_{ap}}{Q_\mu }/2} \right)}}{{{D_\mu }}}
\end{array}
\end{array}
\end{equation}
where \({D_\mu }\), \({D_\nu }\), \({Y_0}'\) and \({E_\mu }'\) respectively denote the overall gain of signal pulses, the weak decoy-state  pulses, the yield of vacuum decoy-state pulses and the observed QBER under after-pulse impacts; \(\eta \), \({p_{dc}}\), \({e_0}\) and \({e_d}\) represent the transmittance of the channel, the background count rate, the background error rate and the probability that a photon hits the wrong detector.

Further, we obtain the secure key rate
\begin{equation}
\begin{array}{l}
\begin{array}{ccl}
\displaystyle R & \ge & q\left\{ {{Q_0} + Q_{_1}^z [ {1 - H\left( {e_1^{pz}} \right)}] - lea{k_{EC}}} \right\}\\
\displaystyle lea{k_{EC}} &=& {f_{EC}}{Q_\mu }H({E_\mu })\\
\displaystyle q &=& \frac{{{N_\mu }{p_z}}}{{{N_{total}}}}\\
\displaystyle {Q_0} &=& {Y_0}'{e^{ - \mu }}
\end{array}
\end{array}
\end{equation}
where \(q\) denotes the received signal state ratio; \({Q_0}\) and \(Q_{_1}^z\) represent the background and the single photon gains, \(e_1^{pz}\) is the phase error rate of single photon signal state; \(lea{k_{EC}}\), \({f_{EC}}\), \({Q_\mu }\)and \({E_\mu }\) separately denote the cost of error correction, the bilateral error correction inefficiency, the overall gain of signal state and quantum bit error rate (QBER); \(H\left( x \right) =  - x{\log _2}\left( x \right) - (1 - x){\log _2}(1 - x)\)is the binary Shannon entropy function;  \({N_{total}}\), \({N_\mu }\), \({p_z}\) and \({Y_0}\) are the length of the sending qubits, the signal pulses in the  basis, the probability that Bob chooses the  basis, and the background count rate.

In security analysis, we need to estimate the lower bound of single photon yield denoted by \({Y_1}\) and the upper bound of its relevant error rate \({e_1}\) measured in the \(Z\) basis. Their expressions are
\begin{equation}
{Y_1} \ge Y_1^L = \frac{\mu }{{\mu v - {v^2}}}\left( {{Q_v}{e^v} - {Q_\mu }{e^\mu }\frac{{{v^2}}}{{{\mu ^2}}} - \frac{{{\mu ^2} - {v^2}}}{{{\mu ^2}}}{Y_0}'} \right)
\end{equation}
and
\begin{equation}
{e_1} \le e_1^U = \frac{{{E_v}{Q_v}{e^v} - {e_0}{Y_0}'}}{{Y_1^Lv}}.
\end{equation}

\subsection{\label{sec:level2B} Mathematical model}
Mathematical model for the corresponding statistical fluctuation analysis considering after-pulse contributions is discussed as follows.
Let \(\overline X \) denote the observed value, which is obtained by detection events of \(m\) random samples denoted by \({{X}_{1}},{{X}_{2}},\cdots ,{{X}_{m}}\), detected with the value 1 and 0 otherwise, satisfying \(\overline{X}=\frac{1}{m}\sum\nolimits_{i=1}^{m}{{{X}_{i}}}\). \(E\) is the expected value of \(\frac{1}{m}\sum\nolimits_{i = 1}^m {{X_i}} \) when \(m\) tends to infinity. Because of after-pulse contributions, we have the following probability relations
\begin{equation}
 \Pr\left\{ {{X_{m + 1}} = s|{X_m} = t,{X_{n - 1}} = {x_{n - 1}}, \cdots ,{X_1} = {x_1},{X_0} = {x_0}} \right\} = Pr\left\{ {{X_{m + 1}} = s|{X_m} = t} \right\} \ne Pr\left\{ {{X_{m + 1}} = s} \right\},
\end{equation}
which shows that the random samples are Markov chain. Then for any   and  , we need to find a proper inequality satisfying
\begin{equation}
 \Pr \left( {\left| {\overline X  - E} \right| \ge \xi } \right) \le \varepsilon \,
\end{equation}
where the parameter   can be seen as a function of  . That is to say the expected value   is contained in the interval
\begin{equation}
\left[ {\overline X  - \xi ,\overline X  + \xi } \right]
\end{equation}
with a failure probability \(\varepsilon \).

\section{\label{sec:level3} Analysis of i.i.d random samples}
The purpose of this section is to make a deviation comparison between two kinds of analysis methods based on i.i.d. random samples.

\subsection{\label{sec:level3A} Standard error analysis}
The original protocol adopts standard error analysis which is the application of central-limit theorem using Gaussian distribution to deal with the statistical fluctuation problems. The application precondition is that the random samples must be i.i.d. and the sum of them has to obey Gaussian distributions. We denote \({\xi _1}({S_{obs}})\) as the deviation when using standard error analysis, here \({S_{obs}}\) refers to the respective random samples. There are four kinds of deviations corresponding to Wei's protocol defined in Appendix~\ref{app:sec1}.

\subsection{\label{sec:level3B} Law of large numbers}
Law of large numbers \cite{Cover2006elements} mainly describes the stable equivalence relation between occurrence frequency and probability distribution when the test number is large enough. Its application precondition only requires that the random samples are i.i.d.. And the mathematical model is stated as follows.

The only difference from the above mentioned one is that the observed value \(\overline X \) is obtained by detection events of \(m\) i.i.d. random samples. Then for any \(\xi  > 0\) and \(\varepsilon  \ge 0\), a quantum version of law of large numbers yields the following statement
\begin{equation}
\begin{array}{ccl}
\Pr \left( {\left| {\overline X  - E} \right| \ge \xi  \equiv \sqrt {\frac{{2\left[ {\ln \left( {\frac{1}{\varepsilon }} \right) + 2\ln (m + 1)} \right]}}{m}} } \right) \le \varepsilon,
\end{array}
\end{equation}
where \(\xi  \equiv \sqrt {\frac{{2\left[ {\ln \left( {\frac{1}{\varepsilon }} \right) + 2\ln (m + 1)} \right]}}{m}} \). Applying law of large numbers in dealing with statistical fluctuation problems in Wei's protocol, we obtain the relevant deviations denoted by \({\xi _2}({S_{obs}})\), which can be found in Appendix~\ref{app:sec2}.

\subsection{\label{sec:level3C} Deviation comparison}
Here we should point out that the quantile denoted by \({u_\alpha }\) in standard error analysis is determined by the failure probability when applying the statistical fluctuation analysis method. Thus, before making a deviation comparison between the two methods, we need to set up a fair comparison standard first. We set a fixed failure probability no matter which method is used. Suppose that the failure probability is \(\varepsilon  = {10^{ - 10}}\), which is equivalent to taking the quantile as \({u_\alpha } = 6.4\). So we have to change the quantile from \({u_\alpha } = 5\) in the original protocol into \({u_\alpha } = 6.4\).

Given that the length of sending qubits is long enough so that the sum of them satisfies Gaussian distribution. Now we just take \({Q_\mu }\) as an example and the respective deviations are \({\xi _1}\left( {{Q_\mu }} \right) = \frac{{6.4}}{{\sqrt {{N_\mu }{p_z}{Q_\mu }} }}\) and \({\xi _2}\left( {{Q_\mu }} \right) = \sqrt {\frac{{2\left[ {\ln {{10}^{10}} + 2\ln ({N_\mu }{p_z} + 1)} \right]}}{{{N_\mu }{p_z}{Q_\mu }^2}}} \).

By some reduction, it equals to compare \(6.4 \cdot \sqrt {{Q_\mu }} \) and \(\sqrt {46.06 + 4\ln ({N_\mu }{p_z} + 1)} \).

Since \(\sqrt {{Q_\mu }}  < 1\) and \(4\ln ({N_\mu }{p_z} + 1) > 0\), then we can obtain the following inequalities
\begin{equation}
\begin{array}{ccl}
\sqrt {46.06 + 4\ln ({N_\mu }{p_z} + 1)}  > \sqrt {46.06}  > 6.4 > 6.4 \cdot \sqrt {{Q_\mu }}.
\end{array}
\end{equation}

As a result, it can be safely concluded that the relationship between relevant deviations satisfies \({\xi _1} < {\xi _2}\). In other words, standard error analysis can achieve a better estimation result than law of large numbers. The simulation result is shown in \ref{sec:level5}.

\section{\label{sec:level4} Analysis of independent random samples}
Large deviation theory \cite{Hoeffding1963probability,chernoff1952measure} describes the remote tails' asymptotic behavior of probability distribution sequences. That is to say, it concerns the exponential decline of the probability measures of certain kinds of tail events. It is a precise form of law of large numbers and the main application precondition only requires the relation of observed samples is independent. Here we take two basic theorems to deal with the statistical fluctuation issues concerning after-pulse contributions.

\subsection{\label{sec:level4A} Hoeffding's inequality}
The theorem introduced first is called Hoeffding's inequality \cite{Hoeffding1963probability}, which provides an upper bound on the probability that the sum of independent random variables deviates from its expected value. A quantum version of Hoeffding's inequality yields the following statement.

Let \(\overline X \) denote the observed value, which is obtained by detection events of \(m\) independent random samples. And \({X_i}\) are almost surely bounded, that is \(\Pr \left( {{X_i} \in \left[ {{a_i},{b_i}} \right]} \right) = 1\), \(1 \le i \le m\).Then for any \(\xi  > 0\) and \(\varepsilon  \ge 0\), the following inequality holds
\begin{equation}
\begin{array}{ccl}
\displaystyle \Pr \left( {\left| {\overline X  - E\left[ {\overline X } \right]} \right| \ge \xi } \right) \le 2{e^{\left( { - \frac{{2{m^2}{\xi ^2}}}{{\sum\nolimits_{i = 1}^m {{{\left( {{b_i} - {a_i}} \right)}^2}} }}} \right)}} = \varepsilon,
\end{array}
\end{equation}
here \(E\left[ {\overline X } \right]\) refers to the expected value of \(\overline X \).

Using Hoeffding's inequality, we get the relevant deviations denoted by \({\xi _3}({S_{obs}})\), which can be found in Appendix~\ref{app:sec1}.

\subsection{\label{sec:level4B} Chernoff bound}
The second theorem introduced here is called Chernoff bound \cite{chernoff1952measure}, also providing an upper bound on the probability that the sum of independent random variables deviates from its expected value. To apply the method in Wei's protocol, we changes the form mentioned in \cite{curty2014finite} into the following claim.

Let \(\overline X \) denote the observed value, which is obtained by detection events of \(m\) independent random samples, satisfying \(\Pr ({X_i} = 1) = {p_i}\), \(\overline X  = \frac{1}{m}\sum\nolimits_{i = 1}^m {{X_i}} \),and \(a = E\left[ {\overline X } \right] = \frac{1}{m}\sum\nolimits_{i = 1}^m {{p_i}} \), where \(E\left[  \cdot  \right]\) denotes the mean value. Let \(x\) be the observed outcome of \(\overline X \) and \({a_L} = x - \sqrt {\frac{1}{{2m}}\ln \left( {\frac{1}{{{\varepsilon _1}}}} \right)} \) for certain \({\varepsilon _1} > 0\). For certain \({\varepsilon _2}\), \({\varepsilon _3} \ge 0\), we have that \(x\) satisfies
\begin{equation}
\begin{array}{ccl}
\displaystyle x = a + \delta,
\end{array}
\end{equation}
except with error probability \(\gamma  = {\varepsilon _1} + {\varepsilon _2} + {\varepsilon _3}\), where \(\delta  \in [ - \Delta ,\Delta ']\). Here \({\varepsilon _2}\)(\({\varepsilon _3}\)) refers to the probability that \(x < a - \Delta \)(\(x > a + \Delta '\)). The limitations and corresponding values of \(\Delta \) and \(\Delta '\) are given in Appendix~\ref{app:sec3}. Employing Chernoff bound in Wei's protocol, we obtain the corresponding deviations denoted by \({\xi _4}({S_{obs}})\) in Appendix~\ref{app:sec1}.

\subsection{\label{sec:level4C} Deviation comparison}
In the definition of Chernoff bound, it can be easily concluded that if the three conditions are not satisfied, then Chernoff bound changes its form into Hoeffding's inequality. Here we only consider the situation that the three conditions in Chernoff bounds are all fulfilled, and the comparisons corresponding to the other results are shown in Appendix~\ref{app:sec2}.

For convenience, we suppose \({\varepsilon _1} = {\varepsilon _2} = {\varepsilon _3} = {10^{ - 10}}\). Still we take   \({Q_\mu }\) as an example and the respective deviations are \({\xi _3}\left( {{Q_\mu }} \right) = \sqrt {\frac{1}{{2{N_\mu }{p_z}}}\ln \frac{1}{\varepsilon }} \), \(\Delta  = \sqrt {\frac{{2{Q_\mu }}}{{{N_\mu }{p_z}}}\ln \frac{{16}}{{{\varepsilon ^4}}}} \) and \(\Delta ' = \sqrt {\frac{{2{Q_\mu }}}{{{N_\mu }{p_z}}}\ln \frac{1}{{{\varepsilon ^{3/2}}}}} \), here \({\xi _3}\), \(\Delta \) and \(\Delta '\) refer to the deviations when using Hoeffding's inequality and Chernoff bound.

Then if both \(\sqrt {\frac{{2{Q_\mu }}}{{{N_\mu }{p_z}}}\ln \frac{{16}}{{{\varepsilon ^4}}}}  \le \sqrt {\frac{1}{{2{N_\mu }{p_z}}}\ln \frac{1}{\varepsilon }} \) and \(\sqrt {\frac{{2{Q_\mu }}}{{{N_\mu }{p_z}}}\ln \frac{1}{{{\varepsilon ^{3/2}}}}}  \le \sqrt {\frac{1}{{2{N_\mu }{p_z}}}\ln \frac{1}{\varepsilon }} \) hold, we can safely say that Chernoff bound can achieves a better result than Hoeffding's inequality.

Simplify the above inequalities, we obtain the sufficient condition is
\begin{equation}
\begin{array}{ccl}
\displaystyle {Q_\mu } \le 0.06.
\end{array}
\end{equation}

\section{\label{sec:level5} Analysis of dependent random samples}
Concerning the worst case that the after-pulse (maybe caused by eavesdropper) exists in every detection event, the mutual relationship between adjacent samples doesn¡¯t satisfy the independent condition anymore. Therefore, we need a stringent way to deal with the problem based on dependent observed samples. Here we try to solve the problem by constructing martingale \cite{grimmett2001brobability} and bring Azuma¡¯s inequality \cite{azuma1967dependent} into statistical fluctuation analysis. Detailed proof is shown in Appendix C. Here the quantum version of Azuma¡¯s inequality is presented as follows.

Let \(\overline X \) denote the observed value, which is obtained by detection events of \(m\) independent random samples, \({M_n} = \frac{1}{n} \cdot \sum\nolimits_{i = 1}^n {{X_i}} \) and  \({M_0} = E\), here \(E\) is the value of \(\frac{1}{n} \cdot \sum\nolimits_{i = 1}^n {{X_i}} \) when \(n\) tends to infinity. Then for any \(\xi > 0\) and \(\varepsilon  \ge 0\), the following inequality holds
\begin{equation}
\begin{array}{ccl}
\displaystyle \Pr \left( {\left| {{M_n} - {M_0}} \right| \ge \xi } \right)&=& \Pr \left( {\left| {\frac{1}{n} \cdot \sum\nolimits_{i = 1}^n {{X_i}}  - E} \right| \ge \xi } \right) \\
\displaystyle & \le & 2{e^{ - \frac{{{n^2}{\xi ^2}}}{{2\sum\nolimits_{i = 1}^n {X_i^2} }}}},
\end{array}
\end{equation}
with \(\xi  = \sqrt {\frac{2}{n}\ln \frac{2}{\varepsilon }} \). Applying Azuma's inequality in solving the statistical fluctuation problems based on dependent samples, we get the relevant deviations denoted by \({\xi _5}({S_{obs}})\) in Appendix~\ref{app:sec1}.

\section{\label{sec:level6} Simulation}
To obtain the impact on secure key rate of after-pulse contribtions, we take Wei's protocol as an example to see the comparisons with and without after-pulse effect.
\begin{figure}[h]
\centering
\includegraphics[height=0.32\textwidth,width=0.5\textwidth]{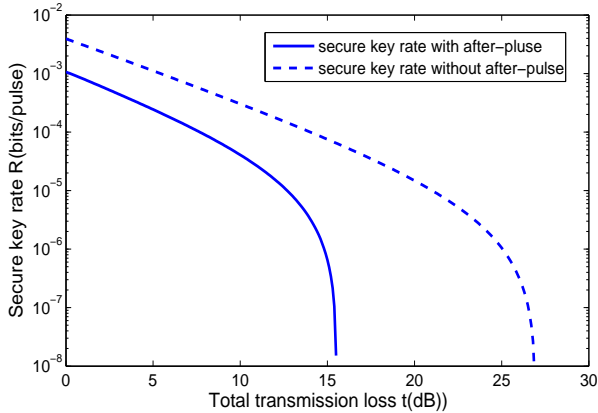}
\caption{\label{fig:keyratecomparison-1}
(Color online) Plot of secure key rate R versus total transmission loss \(t\) with and without after-pulse contributions. The blue solid line denotes the secure key rate under after-pulse impact, and the blue dotted line stands for the secure key rate without after-pulse contributions. Here we take \(N = 6 \times {10^9}\), \({u_\alpha } = 5\).
}
\end{figure}

Figure \ref{fig:keyratecomparison-1} shows that the secure key rate under after-pulse impact is 75\% lower than the one without after-pulse contribution. As a result, the influence of after-pulse on secure key rate can't be neglected.

\begin{figure}[h]
\centering
\includegraphics[height=0.32\textwidth,width=0.5\textwidth]{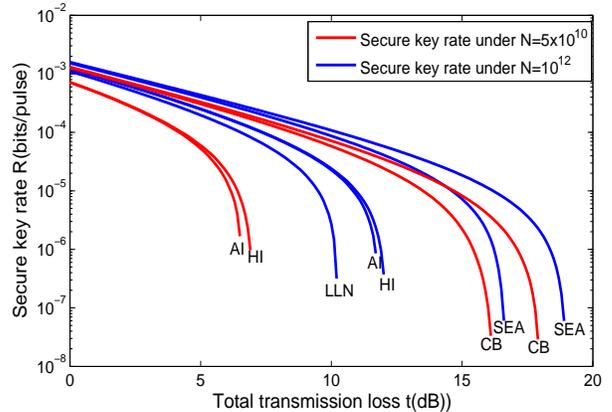}
\caption{\label{fig:keyratecomparison-2}
 (Color online) Plot of secure key rate R versus total transmission loss \(t\) under different estimation methods. The red lines denote the respective secure key rates with \(N = 5 \times {10^{10}}\), and the blue lines refer to the ones with \(N = {10^{12}}\). The abbreviations respectively stands for Law of large numbers (LLN), standard error analysis (SEA), Hoeffding's inequality (HI), Chernoff bound (CB) and Azuma's inequality (AI). Taking \(N = {N_{total}} \in \left\{ {5 \times {{10}^{10}},{{10}^{12}}} \right\}\), \(\varepsilon  = {\varepsilon _1} = {\varepsilon _2} = {\varepsilon _3} = {10^{ - 10}}\), and \({u_\alpha } = 6.4\).
}
\end{figure}

Take the after-pulse contribution into consideration, we obtain the secure key rates under different statistical fluctuation analysis methods in figure \ref{fig:keyratecomparison-2}.

\begin{figure}[h]
\centering
\includegraphics[height=0.32\textwidth,width=0.5\textwidth]{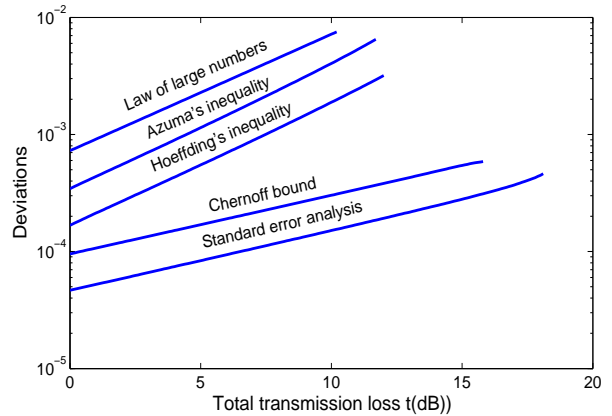}
\caption{\label{fig:deviations}
 (Color online) Plot of deviations comparisons versus total transmission loss \(t\) under various estimation methods. The blue lines from top to bottom respectively stands for the deviations of  Law of large numbers, Azuma's inequality, denotes the deviations when using law of large numbers, Azuma's inequality, Hoeffding's inequality, Hoeffding's inequality , Chernoff bound and standard error analysis with \(N = {10^{12}}\).
}
\end{figure}

And the corresponding deviation comparisons are presented in figure \ref{fig:deviations}.

From the above analysis, we get a clear comparison of the estimation accuracy using different methods. And for different relations of the observed samples, we should choose the proper way to deal with statistical fluctuation problems so as to get a precise result.

\section{\label{sec:level7}Conclusion}
In this paper, we give a detailed analysis in dealing with statistical fluctuation issues, based on independent and dependent random samples. We set up respective mathematical models for each estimation method and give a fair comparison standard, together with the corresponding deviation comparisons. For the first time we put forward a stringent way in handling the statistical fluctuations based on dependent samples. From the result of numerical simulation, we conclude that the after-pulse impact can¡¯t be neglected since it greatly lowers the secure key rate. And if the number of the totthe estimation accuracy of standard error analysis is the best one in the four methods. However, the large deviation theory possesses more generality than standard error analysis. Although the secure key rate using Azuma¡¯s inequality is a little lower than Hoeffding¡¯s inequality, it¡¯s a strict tool dealing with fluctuation problems when taking after-pulse contributions into consideration, and can be extended in other QKD protocols.

We emphasize that Azuma¡¯s inequality is a stringent way in dealing with statistical fluctuations based on dependent samples, but not an efficient one. So there is still more room for accuracy improvement in finding other proper ways.

\appendix

\section{\label{app:sec1}Respective deviations}
Deviations under standard error analysis are represented as follows.
\begin{equation}
\begin{array}{l}
\displaystyle {\xi _1}\left( {{Q_\mu }} \right) = \frac{{{u_\alpha }}}{{\sqrt {{N_\mu }{p_z}{Q_\mu }} }}\\
\displaystyle {\xi _1}\left( {{Q_v}} \right) = \frac{{{Q_v}{u_\alpha }}}{{\sqrt {N_v^z{p_z}{Q_v}} }}\\
\displaystyle {\xi _1}\left( {{Y_0}'} \right) = \frac{{{u_\alpha }}}{{\sqrt {{N_0}{Y_0'}} }}\\
\displaystyle {\xi _1}\left( {{Q_0}} \right) = \frac{{{u_\alpha }}}{{\sqrt {{N_0}{Q_0}} }}
\end{array}
\end{equation}
where \({Q_\mu }\), \({Q_v}\), \({Y_0}\)and \({{Q_0}}\) are the observed values instead of probabilities.

Deviations under Law of large numbers are represented as follows.
\begin{equation}
\begin{array}{l}
\displaystyle Q_\mu ^U = {Q_\mu }\left( {1 + {\xi _2}\left( {{Q_\mu }} \right)} \right) = {Q_\mu } + \sqrt {\frac{{2\left[ {\ln \left( {\frac{1}{\varepsilon }} \right) + 2\ln ({N_\mu }{p_z} + 1)} \right]}}{{{N_\mu }{p_z}}}} \\
\displaystyle Q_v^L = {Q_v}\left( {1 - {\xi _2}\left( {{Q_v}} \right)} \right) = {Q_v} - \sqrt {\frac{{2\left[ {\ln \left( {\frac{1}{\varepsilon }} \right) + 2\ln (N_v^z{p_z} + 1)} \right]}}{{N_v^z{p_z}}}} \\
\displaystyle Y_0^L = {Y_0}'\left( {1 - {\xi _2}\left( {{Y_0}'} \right)} \right) = {Y_0}' - \sqrt {\frac{{2\left[ {\ln \left( {\frac{1}{\varepsilon }} \right) + 2\ln ({N_0} + 1)} \right]}}{{{N_0}}}} \\
\displaystyle Q_0^L = {Q_0}\left( {1 - {\xi _2}\left( {{Q_0}} \right)} \right) = {Q_0} - \sqrt {\frac{{2\left[ {\ln \left( {\frac{1}{\varepsilon }} \right) + 2\ln ({N_0} + 1)} \right]}}{{{N_0}}}}
\end{array}
\end{equation}

Deviations under Hoeffding's inequality are represented as follows.
 \begin{equation}
\begin{array}{l}
\displaystyle Q_\mu ^U = {Q_\mu }\left( {1 + {\xi _3}\left( {{Q_\mu }} \right)} \right) = {Q_\mu } + \sqrt {\frac{1}{{2{N_\mu }{p_z}}}\ln \frac{1}{\varepsilon }} \\
\displaystyle Q_v^L = {Q_v}\left( {1 - {\xi _3}\left( {{Q_v}} \right)} \right) = {Q_v} - \sqrt {\frac{1}{{2N_v^z{p_z}}}\ln \frac{1}{\varepsilon }} \\
\displaystyle Y_0^L = {Y_0}'\left( {1 - {\xi _3}\left( {{Y_0}'} \right)} \right) = {Y_0}' - \sqrt {\frac{1}{{2{N_0}}}\ln \frac{1}{\varepsilon }} \\
\displaystyle Q_0^L = {Q_0}\left( {1 - {\xi _3}\left( {{Q_0}} \right)} \right) = {Q_0} - \sqrt {\frac{1}{{2{N_0}}}\ln \frac{1}{\varepsilon }}
\end{array}
\end{equation}

Deviations under Chernoff bound are represented as follows.
\begin{equation}
\begin{array}{l}
\begin{array}{lcl}
\displaystyle Q_\mu ^U &=& {Q_\mu }\left( {1 + {\xi _4}\left( {{Q_\mu }} \right)} \right) = {Q_\mu } + \sqrt {\frac{{2{Q_\mu }}}{{{N_\mu }{p_z}}}\ln \frac{{16}}{{{\varepsilon _2}^4}}} \\
\displaystyle Q_v^L &=& {Q_v}\left( {1 - {\xi _4}\left( {{Q_v}} \right)} \right)\\
\displaystyle  &=& {Q_v} - \min \left\{ {\sqrt {\frac{1}{{2N_v^z{p_z}}}\ln \frac{1}{{{\varepsilon _1}}}} ,\sqrt {\frac{{2{Q_v}}}{{N_v^z{p_z}}}\ln \frac{1}{{{\varepsilon _3}^{3/2}}}} } \right\}\\
\displaystyle Y_0^L &=& Y{'_0}\left( {1 - {\xi _4}\left( {{Y_0}'} \right)} \right)\\
\displaystyle &=& {Y_0}' - \min \left\{ {\sqrt {\frac{1}{{2{N_0}}}\ln \frac{1}{{{\varepsilon _1}}}} ,\sqrt {\frac{{2{Y_0}'}}{{{N_0}}}\ln \frac{2}{{{\varepsilon _3}^{3/2}}}} } \right\}\\
\displaystyle Q_0^L &=& {Q_0}\left( {1 - {\xi _4}\left( {{Q_0}} \right)} \right)\\
\displaystyle &=& {Q_0} - \min \left\{ {\sqrt {\frac{2}{{2{N_0}}}\ln \frac{1}{{{\varepsilon _1}}}} ,\sqrt {\frac{{2{Q_0}}}{{{N_0}}}\ln \frac{2}{{{\varepsilon _3}^{3/2}}}} } \right\}
\end{array}
\end{array}
\end{equation}
Deviations under Azuma's inequality are represented as follows.
\begin{equation}
\begin{array}{l}
\displaystyle Q_\mu ^U = {Q_\mu }\left( {1 + {\xi _5}\left( {{Q_\mu }} \right)} \right) = {Q_\mu } + \sqrt {\frac{2}{{{N_\mu }{p_z}}}\ln \frac{2}{\varepsilon }} \\
\displaystyle Q_v^L = {Q_v}\left( {1 + {\xi _5}\left( {{Q_v}} \right)} \right) = {Q_v} - \sqrt {\frac{2}{{N_v^z{p_z}}}\ln \frac{2}{\varepsilon }} \\
\displaystyle Y_0^L = Y{'_0}\left( {1 + {\xi _5}\left( {{Y_0}'} \right)} \right) = {Y_0}' - \sqrt {\frac{2}{{{N_0}}}\ln \frac{2}{\varepsilon }} \\
\displaystyle Q_0^L = {Q_0}\left( {1 + {\xi _5}\left( {{Q_0}} \right)} \right) = {Q_0} - \sqrt {\frac{2}{{{N_0}}}\ln \frac{2}{\varepsilon }}
\end{array}
\end{equation}

\section{\label{app:sec2}Chernoff bound}
When it comes to Chernoff bound theory, there are three conditions satisfied or not corresponding to six different results.

i.	\({\left( {2\varepsilon _2^{ - 1}} \right)^{{1 \mathord{\left/
 {\vphantom {1 {m{a_L}}}} \right.
 \kern-\nulldelimiterspace} {m{a_L}}}}} \le {e^{{{({3 \mathord{\left/
 {\vphantom {3 {4\sqrt 2 }}} \right.
 \kern-\nulldelimiterspace} {4\sqrt 2 }})}^2}}}\)

ii.	\({\left( {\varepsilon _3^{ - 1}} \right)^{{1 \mathord{\left/
 {\vphantom {1 {m{a_L}}}} \right.
 \kern-\nulldelimiterspace} {m{a_L}}}}} \le {e^{{1 \mathord{\left/
 {\vphantom {1 3}} \right.
 \kern-\nulldelimiterspace} 3}}}\)

iii.\({\left( {\varepsilon _3^{ - 1}} \right)^{{1 \mathord{\left/
 {\vphantom {1 {m{a_L}}}} \right.
 \kern-\nulldelimiterspace} {m{a_L}}}}} \le {e^{[{{({{2e - 1)} \mathord{\left/
 {\vphantom {{2e - 1)} 2}} \right.
 \kern-\nulldelimiterspace} 2}]}^2}}}\)	

Let \(f\left( {x,y} \right) = \sqrt {\frac{{2x}}{{{m^2}}}\ln \frac{1}{y}} \), then we have the following results.

(1)	If condition (i) and (ii) are fulfilled, then \(\Delta  = f\left( {x,{{\varepsilon _2^4} \mathord{\left/
 {\vphantom {{\varepsilon _2^4} {16}}} \right. \kern-\nulldelimiterspace} {16}}} \right)\) and \(\hat \Delta  = f\left( {x,\varepsilon _3^{3/2}} \right)\).

(2)	If condition (i) and (iii) are fulfilled, while (ii) not, then \(\Delta  = f\left( {x,{{\varepsilon _2^4} \mathord{\left/{\vphantom {{\varepsilon _2^4} {16}}} \right.\kern-\nulldelimiterspace} {16}}} \right)\) and \(\hat \Delta  = f\left( {x,\varepsilon _3^2} \right)\).

(3)	If condition (i) is fulfilled and (iii) not, then \(\Delta  = f\left( {x,{{\varepsilon _2^4} \mathord{\left/
 {\vphantom {{\varepsilon _2^4} {16}}} \right.\kern-\nulldelimiterspace} {16}}} \right)\) and \(\hat \Delta  = f\left( {{m \mathord{\left/{\vphantom {m 4}} \right.\kern-\nulldelimiterspace} 4},\varepsilon } \right)\).

(4)	If condition (ii) is fulfilled and (i) not, then \(\Delta  = f\left( {{m \mathord{\left/
 {\vphantom {m 4}} \right.\kern-\nulldelimiterspace} 4},\varepsilon } \right)\) and \(\hat \Delta  = f\left( {x,\varepsilon _3^{3/2}} \right)\).

(5)	If condition (i) and (ii) are not fulfilled, while (iii) is, then \(\Delta  = f\left( {{m \mathord{\left/
 {\vphantom {m 4}} \right.\kern-\nulldelimiterspace} 4},\varepsilon } \right)\) and \(\hat \Delta  = f\left( {x,\varepsilon _3^2} \right)\).

(6)	If condition (i) , (ii) and  (iii) are not fulfilled then \(\Delta  = \hat \Delta  = f\left( {{m \mathord{\left/
 {\vphantom {m 4}} \right.\kern-\nulldelimiterspace} 4},\varepsilon } \right)\).

The other deviation comparison results with Hoeffding's inequality are stated here, also taking \({Q_\mu }\) as an example and \({\varepsilon _1} = {\varepsilon _2} = {\varepsilon _3} = {10^{ - 10}}\).

(a) Deviation comparison between  \(\sqrt {\frac{{2{Q_\mu }}}{{{N_\mu }{p_z}}}\ln \frac{{16}}{{{\varepsilon ^4}}}} \), \(\sqrt {\frac{{2{Q_\mu }}}{{{N_\mu }{p_z}}}\ln \frac{1}{{{\varepsilon ^2}}}} \) and \(\sqrt {\frac{1}{{2{N_\mu }{p_z}}}\ln \frac{1}{\varepsilon }} \), under result (2).

The sufficient condition that Chernoff bound achieves a better result than Hoeffding's inequality is
\begin{equation}
\begin{array}{l}
\left\{ {\begin{array}{*{20}{c}}
\displaystyle {\sqrt {\frac{{2{Q_\mu }}}{{{N_\mu }{p_z}}}\ln \frac{{16}}{{{\varepsilon ^4}}}}  \le \sqrt {\frac{1}{{2{N_\mu }{p_z}}}\ln \frac{1}{\varepsilon }} }\\
\displaystyle {\sqrt {\frac{{2{Q_\mu }}}{{{N_\mu }{p_z}}}\ln \frac{1}{{{\varepsilon ^2}}}}  \le \sqrt {\frac{1}{{2{N_\mu }{p_z}}}\ln \frac{1}{\varepsilon }} }
\end{array}} \right.
\end{array}
\end{equation}
This condition can be equivalently written as
\begin{equation}
\begin{array}{l}
{Q_\mu } \le 0.06.
\end{array}
\end{equation}

(b) Deviation comparison between \(\sqrt {\frac{{2{Q_\mu }}}{{{N_\mu }{p_z}}}\ln \frac{{16}}{{{\varepsilon ^4}}}} \) and \(\sqrt {\frac{1}{{2{N_\mu }{p_z}}}\ln \frac{1}{\varepsilon }} \), under result (3).

Obviously, the condition is the same as (a).

(c) Deviation comparison between \(\sqrt {\frac{{2{Q_\mu }}}{{{N_\mu }{p_z}}}\ln \frac{1}{{{\varepsilon ^{{\raise0.7ex\hbox{$3$} \!\mathord{\left/{\vphantom {3 2}}\right.\kern-\nulldelimiterspace}
\!\lower0.7ex\hbox{$2$}}}}}}} \) and \(\sqrt {\frac{1}{{2{N_\mu }{p_z}}}\ln \frac{1}{\varepsilon }} \), under result (4).

Here the sufficient condition is
\begin{equation}
\begin{array}{l}
\displaystyle \sqrt {\frac{{2{Q_\mu }}}{{{N_\mu }{p_z}}}\ln \frac{1}{{{\varepsilon ^{{\raise0.7ex\hbox{$3$} \!\mathord{\left/
 {\vphantom {3 2}}\right.\kern-\nulldelimiterspace}
\!\lower0.7ex\hbox{$2$}}}}}}}  \le \sqrt {\frac{1}{{2{N_\mu }{p_z}}}\ln \frac{1}{\varepsilon }}
\end{array}
\end{equation}
And we have
\begin{equation}
\begin{array}{l}
{Q_\mu } \le 0.162.
\end{array}
\end{equation}

(d) Deviation comparison between \(\sqrt {\frac{{2{Q_\mu }}}{{{N_\mu }{p_z}}}\ln \frac{1}{{{\varepsilon ^2}}}} \) and  \(\sqrt {\frac{1}{{2{N_\mu }{p_z}}}\ln \frac{1}{\varepsilon }} \), under result (5).

Here the sufficient condition is
\begin{equation}
\begin{array}{l}
\displaystyle \sqrt {\frac{{2{Q_\mu }}}{{{N_\mu }{p_z}}}\ln \frac{1}{{{\varepsilon ^2}}}}  \le \sqrt {\frac{1}{{2{N_\mu }{p_z}}}\ln \frac{1}{\varepsilon }}
\end{array}
\end{equation}
And we have
\begin{equation}
\begin{array}{l}
{Q_\mu } \le 0.122.
\end{array}
\end{equation}

(e) Chernoff bound changes its form into Hoeffding's inequality under result (6).

\section{\label{app:sec3}Detailed proof}
First let's look at the definition of Azuma's inequality.

\textit{\textbf{Definition:}}Let \(\left\{ {{Y_1},{Y_2}, \cdots ,{Y_n}} \right\}\) be a martingale, and suppose that there exists a sequence \(\left\{ {{X_1},{X_2}, \cdots ,{X_n}} \right\}\) of real numbers such that \(\Pr \left( {\left| {{Y_n} - {Y_{n - 1}}} \right| \le {K_n}} \right) = 1\) for all \(n\). Then
\begin{equation}
\begin{array}{l}
\displaystyle \Pr \left( {\left| {{Y_n} - {Y_0}} \right| \ge \xi } \right) \le 2{e^{( - \frac{1}{2}{\xi ^2}/\sum\nolimits_{i = 1}^n {X_i^2} )}},\xi  > 0.
\end{array}
\end{equation}

Next we construct the martingale through analyzing the mutual relationships between dependent samples. If the observed value \(\overline X \) are obtained by detection events of \(m\) dependent random samples denoted by \({X_1},{X_2}, \cdots ,{X_m}\), detected with the value 1 and 0 otherwise, satisfying \(\overline X  = \frac{1}{m}\sum\nolimits_{i = 1}^m {{X_i}} \). Considering after-pulse contributions we have the following probability relations
\begin{equation}
\begin{array}{l}
\begin{array}{ccl}
\displaystyle & & Pr\left\{ {{X_{m + 1}} = s|{X_m} = t,{X_{n - 1}} = {x_{n - 1,}} \cdots ,{X_0} = {x_0}} \right\} \\
\displaystyle &=&Pr\left\{ {{X_{m + 1}} = s|{X_m} = t} \right\}\\
\displaystyle &\ne & Pr\left\{ {{X_{m + 1}} = s} \right\}
\end{array}
\end{array}
\end{equation}
(D2) shows that the relationship of the samples satisfies a Markov chain. And take signal pulse as an example, we have \(P\left\{ {{X_{m + 1}} = 1} \right\} = {Q_{{\mu _m}}}\) and \(P\left\{ {{X_{m + 1}} = 1\left| {{X_m} = t} \right.} \right\} = {D_{{\mu _m}}}\), here \({Q_{{\mu _m}}}\) and \({D_{{\mu _m}}}\) are the measured values of overall signal state gain with and without considering after-pulse contributions, when the number of observed samples is \(m\), satisfying \({D_{{\mu _m}}} = {Q_{{\mu _m}}}\left( {1 + {p_{ap}}} \right)\). Let \({S_n} = \sum\nolimits_{i = 1}^n {{X_i}}  = k\), then \({D_{{\mu _n}}} = \frac{{{S_n}}}{n} = \frac{k}{n}\). And we have the following relationship.
\begin{equation}
\begin{array}{l}
\begin{array}{ccl}
\displaystyle & & E({S_{n + 1}}\left| {{S_n} = k} \right.)\\
\displaystyle  &=& k \cdot P({S_{n + 1}} = k\left| {{S_n} = k} \right.) \\
\displaystyle & & + (k + 1) \cdot P({S_{n + 1}} = k + 1\left| {{S_n} = k} \right.)\\
\displaystyle &=& k \cdot P\left\{ {{X_{m + 1}} = 0\left| {{X_m} = t} \right.} \right\}\\
\displaystyle & & + (k + 1) \cdot P\left\{ {{X_{m + 1}} = 1\left| {{X_m} = t} \right.} \right\}\\
\displaystyle  &=& k \cdot \left( {1 - {D_{{\mu _n}}}} \right) + (k + 1) \cdot {D_{{\mu _n}}}\\
\displaystyle  &=& k + {D_{{\mu _n}}}\\
\displaystyle  &=& {S_n} + {D_{{\mu _n}}}
\end{array}
\end{array}
\end{equation}

Let \({M_n} = \frac{{{S_n}}}{n}\),\({M_0} = E\) ,  here \(E\) is the value of \(\frac{1}{n} \cdot \sum\nolimits_{i = 1}^n {{X_i}} \) when \(n\) tends to infinity. We have
\begin{equation}
\begin{array}{l}
\begin{array}{ccl}
\displaystyle  E({M_{n + 1}}\left| {{S_1}, \cdots } \right.,{S_n}) &=& E({M_{n + 1}}\left| {{S_n}} \right.)\\
\displaystyle   &=& E(\frac{{{S_{n + 1}}}}{{n + 1}}\left| {{S_n}} \right.)\\
\displaystyle   &=& \frac{1}{{n + 1}}E({S_{n + 1}}\left| {{S_n}} \right.)\\
\displaystyle   &=& \frac{1}{{n + 1}}({S_n} + {D_{{\mu _n}}})\\
\displaystyle   &=& \frac{{{S_n} + {S_n}/n}}{{n + 1}}\\
\displaystyle   &=& \frac{{{S_n}}}{n} = {M_n}
\end{array}
\end{array}
\end{equation}

From the above analysis we can conclude that \({M_n}\) is a martingale which can be used in Azuma's inequality.

\section*{Acknowledgements}
This work is supported by the National High Technology Research and Development Program of China
Grant No.2011AA010803, the National Natural Science Foundation of China Grants No.61472446 and No.U1204602 and the Open
Project Program of the State Key Laboratory of Mathematical Engineering and Advanced Computing Grant No.2013A14.

\bibliography{fluctuation}

\begin{thebibliography}{10}

\bibitem{bennett1984quantum}
Charles~H Bennett, Gilles Brassard, et~al.
\newblock Quantum cryptography: Public key distribution and coin tossing.
\newblock In {\em Proceedings of IEEE International Conference on Computers,
  Systems and Signal Processing}, volume 175. New York, 1984.

\bibitem{ekert1991quantum}
Artur~K Ekert.
\newblock Quantum cryptography based on bell¡¯s theorem.
\newblock {\em Physical review letters}, 67(6):661, 1991.

\bibitem{gisin2002quantum}
Nicolas Gisin, Gr{\'e}goire Ribordy, Wolfgang Tittel, et~al.
\newblock Quantum cryptography.
\newblock {\em Reviews of Modern Physics}, 74(1):145--190, 2002.

\bibitem{mayers2001unconditional}
Dominic Mayers.
\newblock Unconditional security in quantum cryptography.
\newblock {\em Journal of the ACM (JACM)}, 48(3):351--406, 2001.

\bibitem{bennett1996limitations}
Charles~H Bennett, David~P DiVincenzo, John~A Smolin, and William~K Wootters.
\newblock Mixed-state entanglement and quantum error correction.
\newblock {\em Physical Review A}, 54(5):3824, 1996.

\bibitem{lo1999unconditional}
Hoi-Kwong Lo and H~F Chau.
\newblock Unconditional security of quantum key distribution over arbitrarily
  long distances.
\newblock {\em Science}, 283(5410):2050, 1999.

\bibitem{shor2000simple}
Peter~W Shor and John Preskill.
\newblock Simple proof of security of the bb84 quantum key distribution
  protocol.
\newblock {\em Physical Review Letters}, 85(2):441, 2000.

\bibitem{renner2005security}
Renato Renner.
\newblock {\em Security of quantum key distribution}.
\newblock Ph.d.thesis, Swiss Federal Institute of Technology Zurich, September
  2005.

\bibitem{GLLP2004security}
Daniel Gottesman, Hoi-Kwong Lo, Norbert L{\"u}tkenhaus, and John Preskill.
\newblock Security of quantum key distribution with imperfect devices.
\newblock In {\em Information Theory, 2004. ISIT 2004. Proceedings.
  International Symposium on}, page 136. IEEE.

\bibitem{hwang2003quantum}
Won-Young Hwang.
\newblock Quantum key distribution with high loss: Toward global secure
  communication.
\newblock {\em Physical Review Letters}, 91(5):057901, 2003.

\bibitem{wang2005beating}
Xiang-Bin Wang.
\newblock Beating the photon-number-splitting attack in practical quantum
  cryptography.
\newblock {\em Physical review letters}, 94(23):230503, 2005.

\bibitem{lo2005decoy}
Hoi-Kwong Lo, Xiongfeng Ma, and Kai Chen.
\newblock Decoy state quantum key distribution.
\newblock {\em Physical Review Letters}, 94(23):230504, 2005.

\bibitem{ma2005security}
Xiongfeng Ma.
\newblock Security of quantum key distribution with realistic devices.
\newblock {\em arXiv preprint quant-ph/0503057}, 2005.

\bibitem{brassard2000limitations}
Gilles Brassard, Norbert L{\"u}tkenhaus, Tal Mor, and Barry~C Sanders.
\newblock Limitations on practical quantum cryptography.
\newblock {\em Physical Review Letters}, 85(6):1330, 2000.

\bibitem{pns2002quantum}
Norbert L{\"u}tkenhaus and Mika Jahma.
\newblock Quantum key distribution with realistic states: photon-number
  statistics in the photon-number splitting attack.
\newblock {\em New Journal of Physics}, 4(1):44, 2002.

\bibitem{hayashi2007upper}
Masahito Hayashi.
\newblock Upper bounds of eavesdropper¡¯s performances in finite-length code
  with the decoy method.
\newblock {\em Physical Review A}, 76(1):012329, 2007.

\bibitem{scarani2008quantum}
Valerio Scarani and Renato Renner.
\newblock Quantum cryptography with finite resources: Unconditional security
  bound for discrete-variable protocols with one-way postprocessing.
\newblock {\em Physical review letters}, 100(20):200501, 2008.

\bibitem{cai2009finite}
Raymond~YQ Cai and Valerio Scarani.
\newblock Finite-key analysis for practical implementations of quantum key
  distribution.
\newblock {\em New Journal of Physics}, 11(4):045024, 2009.

\bibitem{tomamichel2012tight}
Marco Tomamichel, Charles Ci~Wen Lim, Nicolas Gisin, and Renato Renner.
\newblock Tight finite-key analysis for quantum cryptography.
\newblock {\em Nature communications}, 3:634, 2012.

\bibitem{ma2012statistical}
Xiongfeng Ma, Chi-Hang~Fred Fung, and Mohsen Razavi.
\newblock Statistical fluctuation analysis for measurement-device-independent
  quantum key distribution.
\newblock {\em Physical Review A}, 86(5):052305, 2012.

\bibitem{wei2013decoy}
Zheng-Chao Wei, Wei-Long Wang, Zhen Zhang, Ming Gao, Zhi Ma, and Xiong-Feng Ma.
\newblock Decoy-state quantum key distribution with biased basis choice.
\newblock {\em Scientific reports}, 3, 2013.

\bibitem{zhou2014tightened}
Yi-Heng Zhou, Zong-Wen Yu, and Xiang-Bin Wang.
\newblock Tightened estimation can improve the key rate of
  measurement-device-independent quantum key distribution by more than 100\%.
\newblock {\em Physical Review A}, 89(5):052325, 2014.

\bibitem{zhou2015making}
Yi-Heng Zhou, Zong-Wen Yu, and Xiang-Bin Wang.
\newblock Making the decoy-state measurement-device-independent quantum key
  distribution practically useful.
\newblock {\em arXiv preprint quant-ph/1502.01262v2}, 2015.

\bibitem{lucamarini2013efficient}
M~Lucamarini, KA~Patel, JF~Dynes, B~Fr{\"o}hlich, AW~Sharpe, AR~Dixon, ZL~Yuan,
  RV~Penty, and AJ~Shields.
\newblock Efficient decoy-state quantum key distribution with quantified
  security.
\newblock {\em Optics express}, 21(21):24550--24565, 2013.

\bibitem{lim2014concise}
Charles Ci~Wen Lim, Marcos Curty, Nino Walenta, Feihu Xu, and Hugo Zbinden.
\newblock Concise security bounds for practical decoy-state quantum key
  distribution.
\newblock {\em Physical Review A}, 89(2):022307, 2014.

\bibitem{curty2014finite}
Marcos Curty, Feihu Xu, Wei Cui, Charles Ci~Wen Lim, Kiyoshi Tamaki, and
  Hoi-Kwong Lo.
\newblock Finite-key analysis for measurement-device-independent quantum key
  distribution.
\newblock {\em Nature communications}, 5, 2014.

\bibitem{lo2008efficient}
Hoi-Kwong Lo, H~F Chau, and Ardehali M.
\newblock Efficient quantum key distribution scheme and a proof of its
  unconditional security.
\newblock {\em Journal of CRYPTOLOGY}, 18(2):133--165, 2008.

\bibitem{tang2014mdi}
Yan-Lin Tang, Hua-Lei Yin, Si-Jing Chen, et~al.
\newblock Measurement-device-independent quantum key distribution over 200 km.
\newblock {\em Physical Review Letters}, 113(7):190501, 2014.

\bibitem{canetti2001uc}
Ran Canetti.
\newblock Universally composable security: A new paradigm for cryptographic
  protocols.
\newblock In {\em IEEE Symposium on Foundations of Computer Science}, volume
  136. New York, 2001.

\bibitem{Cover2006elements}
Thomas~M Cover and Joy~A Thomas.
\newblock {\em Elements of Information Theory}.
\newblock Wiley Series in Telecommunications and Signal Processing.
  Wiley-Interscience, New Jersey, second edition, July 2006.

\bibitem{clopper1934confidence}
C~Clopper and E~S Pearson.
\newblock The use of confidence or fiducial limits illustrated in the case of
  the binomial.
\newblock {\em Biometrika}, 26(4):404--413, 1934.

\bibitem{Hoeffding1963probability}
W~Hoeffding.
\newblock Probability inequalities for sums of bounded random variables.
\newblock {\em Journal of the American Statistical Association},
  58(301):13--30, 1963.

\bibitem{chernoff1952measure}
Herman Chernoff.
\newblock A measure of asymptotic effciency for tests of a hypothesis based on
  the sum of observations.
\newblock {\em Annals of Mathematical Statistics}, 23(4):493--507, 1952.

\bibitem{grimmett2001brobability}
Geoffrey~R Grimmett and Stirzaker~David R.
\newblock {\em Probability and Random Processes}.
\newblock Oxford university press, New York, third edition, April 2001.

\bibitem{azuma1967dependent}
Kazuoki Azuma.
\newblock Weighted sums of certain dependent random variables.
\newblock {\em Tohoku Mathematical Journal}, 19(1967):357--367, 1967.

\bibitem{renner2007symmetry}
Renato Renner.
\newblock Symmetry of large physical systems implies independence of
  subsystems.
\newblock {\em Nature physics}, 3:645--649, 2007.

\bibitem{nambu2011afterpulse}
Nambu Y., Takahashi S., Yoshino K., et~al.
\newblock Efficient and low-noise single-photon avalanche photodiode for
  1.244-ghz clocked quantum key distribution.
\newblock {\em Optics express}, 19(21):20531, 2011.

\bibitem{nato2009afterpulse}
Namekata Nato, Adachi Shunsuke, and Inoue Shuichiro.
\newblock 1.5 ghz single-photon detection at telecommunication wavelengths
  using sinusoidally gated ingaas/inp avalanche photodiode.
\newblock {\em Optics express}, 17(8):6275, 2009.

\bibitem{rolando2013pns}
Rolando~D Somma and Richard~J Hughes.
\newblock Security of decoy-state protocols for general photon-number-splitting
  attacks.
\newblock {\em Physical Review A}, 87(6):062330, 2013.

\bibitem{akio2002method}
Akio Yoshizawa, Ryosaku Kaji, and Hidemi Tsuchida.
\newblock A method of discarding after-pulses in single-photon detection for
  quantum key distribution.
\newblock {\em Japanese Journal of Applied Physics}, 41(10):6016--6017, 2013.

\end{thebibliography}
\bibliographystyle{unsrt}
\end{document}